\documentclass{iau} 
\usepackage{graphicx}
\usepackage{subfigure}

\title[Breaking the disc-halo degeneracy] 
{Breaking the disc-halo degeneracy \\ in NGC 1291 using hydrodynamic simulations}

\author[Francesca Fragkoudi, E. Athanassoula \& A. Bosma]   
{Francesca Fragkoudi$^{1,2}$,
 E. Athanassoula$^1$ \and A. Bosma$^1$}

\affiliation{$^1$Aix Marseille Universit\'{e}, CNRS, LAM (Laboratoire d'Astrophysique de Marseille) UMR 7326, 13388, Marseille, France \\[\affilskip]
$^2$ GEPI, Observatoire de Paris, CNRS, Univ Paris Diderot, Sorbonne Paris Cite, Place Jules Janssen, 92195 Meudon, France \\email: {\tt francesca.fragkoudi@obspm.fr}
}

\pubyear{2016}
\volume{321}  
\jname{Outskirts of Galaxies}
\editors{A. Gil de Paz, J. Knapen \& J. Lee, eds.}
\begin{document}

\maketitle

\begin{abstract}
We present a pilot study on the nearby massive galaxy NGC~1291, in which we aim to constrain the dark matter in the inner regions, by obtaining a dynamical determination of the disc mass-to-light ratio (M/L). To this aim, we model the bar-induced dust lanes in the galaxy, using hydrodynamic gas response simulations. The models have three free parameters, the M/L of the disc, the bar pattern speed and the disc height function. We explore the parameter space to find the best fit models, i.e. those in which the morphology of the shocks in the gas simulations matches the observed dust lanes.
The best-fit models suggest that the M/L of NGC~1291 agrees with that predicted by stellar population synthesis models in the near-infrared ($\approx$0.6\,$M_{\odot}/L_{\odot}$), which leads to a borderline maximum disc for this galaxy. The bar rotates fast, with corotation radius $\leq$ 1.4 times the bar length. Additionally, we find that the height function has a significant effect on the results, and can bias them towards lower or higher M/L.

\end{abstract}

\firstsection 
\section{Introduction}
\begin{figure}[b]
\begin{center}
\subfigure[Comparison]{%
	\includegraphics[height=5cm]{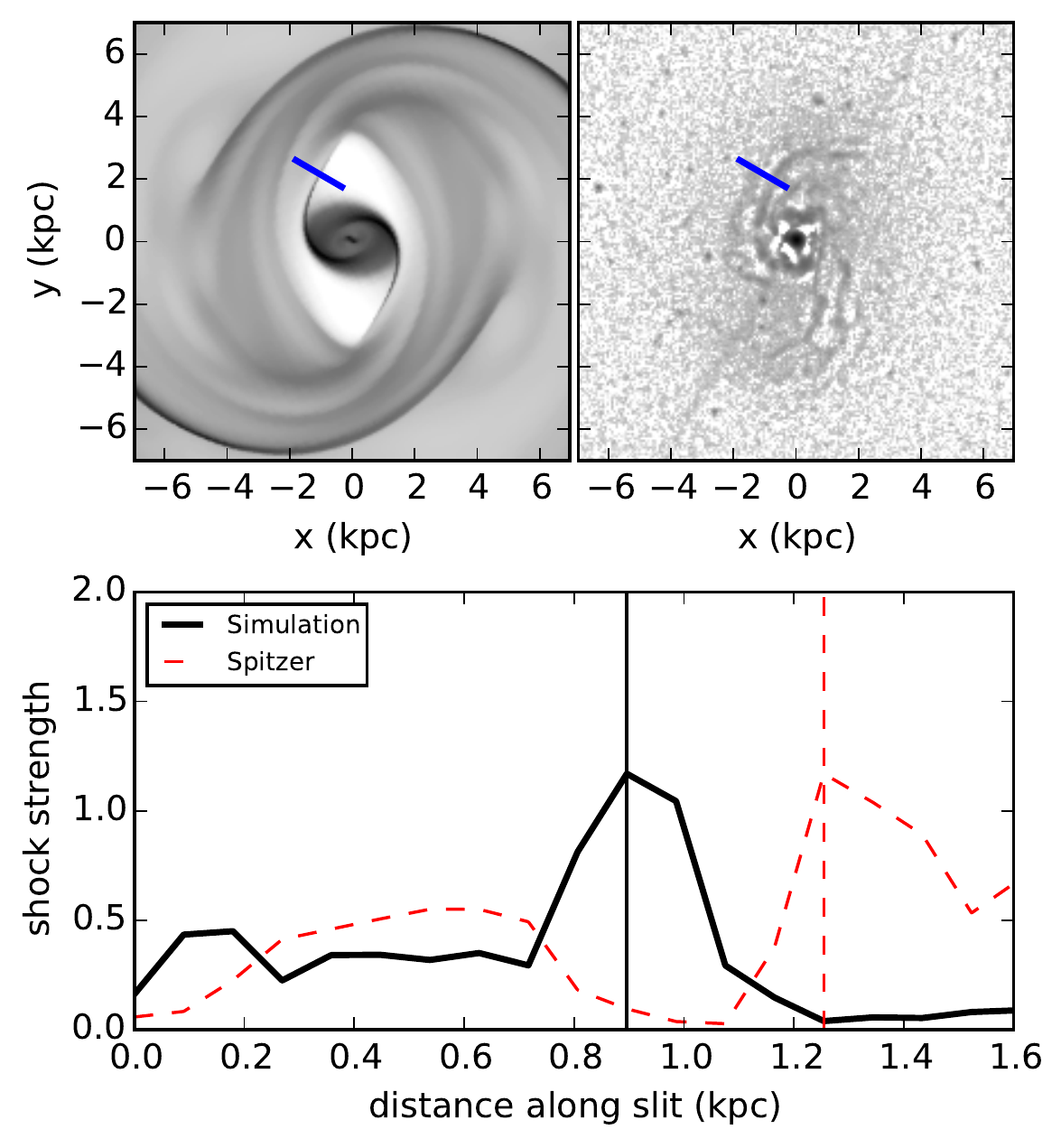} 
	\label{fig:mod005_slit}}
\quad
\subfigure[Rotation curve for different M/L]{%
	\includegraphics[height=5cm]{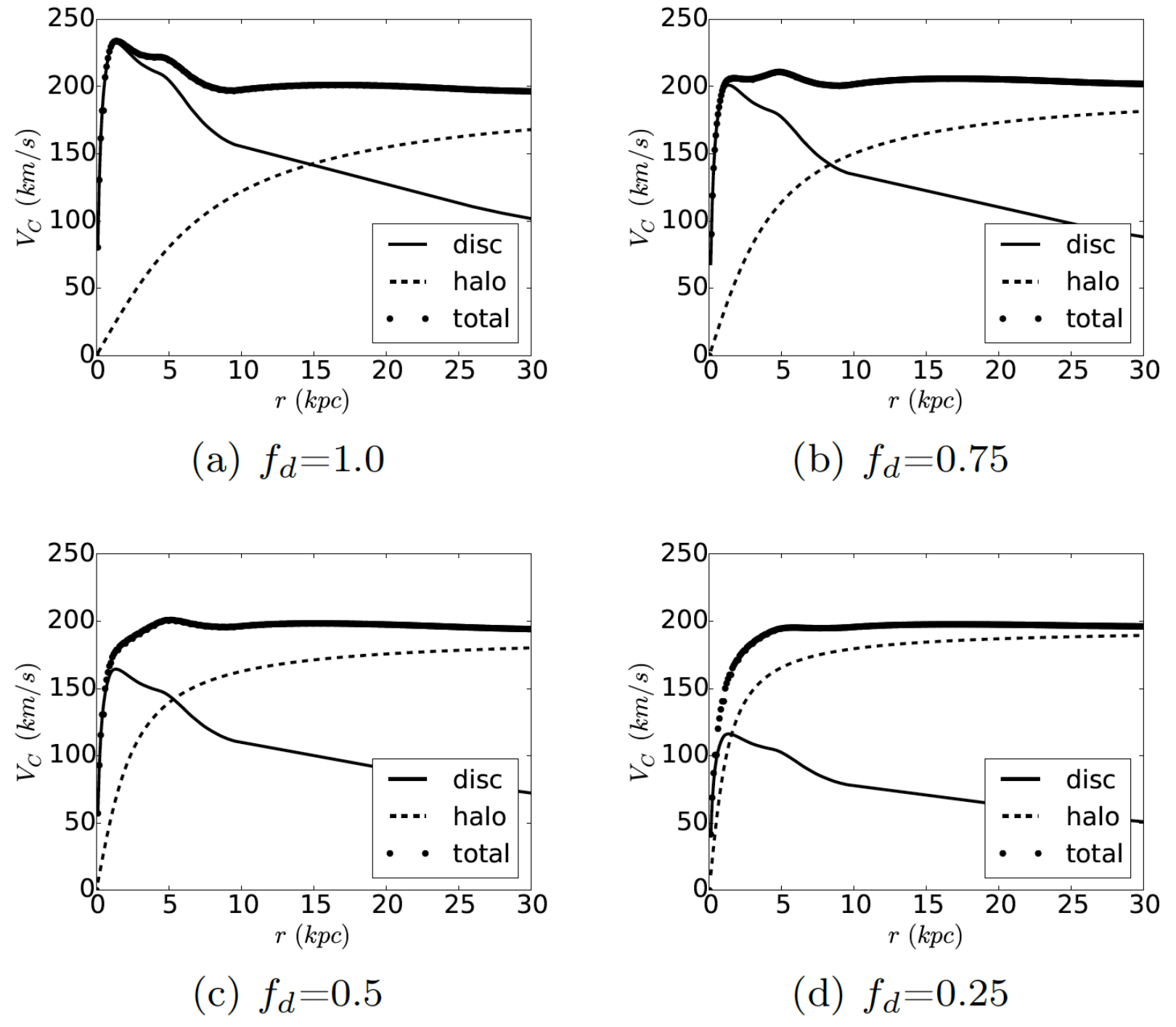}
	\label{fig:rots}}
\quad
\caption{\emph{(a)} \emph{Top:} Comparison between the location of the shocks in the models (left) and the observations (right). A pseudo-slit is placed perpendicular to the shock (thick diagonal line). \emph{Bottom:} The density along the slit in the model (solid black line) and in the observations (dashed red line) as a function of distance along the slit. The shock locus in the model (solid vertical line) and in the observations (dashed vertical line) are obtained from the loci of maximum density. The goodness of fit, ``$\Delta l$'', is defined by the distance between the shock loci. \emph{(b)} The rotation curves for four models with different M/L, where all the other parameters are maintained constant. We see that when we decrease the M/L we need to increase the halo concentration in order to maintain the rotation curve flat at the value predicted by the Tully-Fisher relation.}
\end{center}
\end{figure}

The distribution of dark matter in galaxies has been a source of debate in the scientific community, with one of the main sources of uncertainty being the ``disc-halo degeneracy'' (\cite[van Albada et al. 1985]{vanAlbada1985}). This degeneracy arises because rotation curve decompositions depend critically on the mass-to-light ratio (M/L or $\Upsilon$) which is assigned to the stellar disc component. Rotation curves can be fit equally well with barely any disc contribution, or where the disc contributes the maximum amount possible -- also known as the ``maximum disc hypothesis'' (\cite[van Albada et al. 1985]{vanAlbada1985}, \cite[Sackett 1997]{Sackett1997}). There is no conclusive evidence to either prove or disprove the maximum disc hypothesis and in fact there are a number of arguments both for and against it for the Milky-Way (e.g. \cite[Sackett 2003]{Sackett1997}) as well as for external galaxies (e.g. \cite[Bottema 1993]{Bottema1993}, \cite[Weiner et al. 2001]{Weiner2001}, \cite[Kranz et al. 2003]{Kranzetal2003}).

A way to break the degeneracy between the disc and the dark matter halo is by obtaining dynamical estimates of the M/L ratio of the disc. One way to do this is via the non-axisymmetries induced by the bar, which shocks the gas and leads to the well known ``dust lanes'' observed in barred galaxies (\cite[Prendergast 1983]{Prendergast1983}). These shocks can be reproduced in hydrodynamic simulations (\cite[Athanassoula 1992]{Athanassoula1992}), and their morphology is dependent on the M/L of the disc, the bar pattern speed and the height function of the disc. 

In this study we examine the dust lanes present in the nearby galaxy NGC~1291, and aim to reproduce their morphology using hydrodynamic gas response simulations. The major assumption of this method for obtaining the M/L is that the halo is axisymmetric, and that all the non-axisymmetry in the potential is due to the stellar component. 

\section{Comparing the models to the observations}

There are three main free parameters in the models, and we vary these parameters in order to explore the allowed parameter space. This leads to more than 300 simulations, from which the best fit models are obtained according to how well the shock loci in the models match those in the observations. This is done by placing pseudo-slits perpendicular to the shocks, as shown in Figure \ref{fig:mod005_slit}. The three free parameters are varied as follows:

{\underline{\it M/L}}. To obtain the potential we use a 3.6$\mu m$ image of NGC~1291 from the S$^4$G survey (\cite[Sheth et al. 2010]{Shethetal2010}) to which we apply a M/L. We assume a fiducial value of $\Upsilon_{3.6}$=0.6$M_{\odot}/L_{\odot}$ for the 3.6$\mu$m band (\cite[Meidt et al. 2014]{Meidtetal2014}, \cite[Roeck et al. 2015]{Roecketal2015}), colour corrected to account for the presence of dust by combining information from the 3.6 and 4.5$\mu$m images (\cite[Eskew et al. 2012]{Eskewetal2012}, \cite[Querejeta et al. 2015]{Querejetaetal2015}). We then vary the M/L from 1.5 to 0.25 $\Upsilon_{3.6}$ in steps of 0.25 and add a dark matter halo which maintains the total rotation curve flat at a value of 220\,km/s (see Figure \ref{fig:rots}) as found from the Baryonic Tully-Fisher relation (\cite[McGaugh et al. 2000]{McGaughetal2000}).

{\underline{\it Pattern speed (Lagrangian radius)}}. By varying the Lagrangian radius while keeping all the other parameters of a model constant, we effectively vary the pattern speed of the bar. The Lagrangian radius is varied between 1 and 2 times the bar semi-major axis, in steps of 0.2.

{\underline{\it Height function and scaleheight}}. We vary the height function of the disc by assuming two functional forms -- isothermal and exponential -- and we also vary the scaleheight from $z_0$=0.5 to $z_0$=1.5\,kpc.

\section{Results}

{\underline{\it Dynamical M/L}}: The best fit models obtained in this study have a M/L in the range of M/L$_{3.6}$=0.6-0.75 which falls within the range predicted by stellar population synthesis models and other approaches (i.e. $\Upsilon_{3.6}$=0.6M$_\odot$/L$_\odot$ by e.g. \cite[Meidt et al. (2014)]{Meidtetal2014} and \cite[Roeck et al. (2015)]{Roecketal2015}). It is important to note that our results are independent of systematic uncertainties such as the choice of IMF.

{\underline{\it The maximal disc of NGC 1291}}: The best fit models have borderline a maximal disc on average, with the disc contributing (on average) 74\% of the total rotation velocity. It is worth noting that we are using the definition of Sackett (1997) for disc maximality, which examines the rotation velocity at 2.2$h_r$; in this context, and because the scalelength of NGC~1291 is unusually large (5.8\,kpc), we are examining the rotation velocity at a radius of almost 13\,kpc. Therefore, even for cases where the disc is not according to this definition ``maximal'', the stellar disc dominates the rotation curve in the inner regions of the galaxy.

{\underline{\it The height function of the disc}}: It's furthermore important to note that for this galaxy, models with exponential scaleheights lead to sub maximal discs, while models with an isothermal scaleheight lead to maximal discs. This indicates that the choice of height function, especially for such borderline cases, can affect whether we obtain a maximal disc or not. Additionally, the scaleheight affects the results, as was already shown in previous works, such as in \cite[Fragkoudi et al. (2015)]{Fragkoudietal2015}, where we showed the important effect a boxy/peanut height function can have on the models. 

{\underline{\it Fast rotating bar}}: The best fit models have fast rotating bars with $\mathcal{R}$ $\leq$ 1.4. This is consistent with observational estimates of the bar pattern speed for early type galaxies (\cite[Elmegreen 1996]{Elmegreen1996} and \cite[Corsini 2011]{Corsini2011}). The results are also in agreement with theoretical values (\cite[Athanassoula 1992b]{Athanassoula1992b}).

\end{document}